\documentclass[10pt,conference,letterpaper]{IEEEtran}

\usepackage{amsmath,amssymb}
\usepackage{newtxtext,newtxmath}
\usepackage{graphicx}
\usepackage{cite}
\usepackage{url}
\usepackage[table]{xcolor}
\usepackage{xspace}
\usepackage{booktabs}
\usepackage{multirow}
\usepackage{siunitx}
\usepackage{tikz}
\usepackage{pgfplots}
\usepackage[hidelinks]{hyperref}

\usetikzlibrary{arrows.meta,calc,fit,patterns,positioning,shapes.geometric}
\usepgfplotslibrary{groupplots,colormaps}
\pgfplotsset{compat=1.18}
\newcommand{\system}{\textsc{RAC}\xspace}
\newcommand{\ltoC}{H_{\mathrm{l2c}}}
\newcommand{\ctoL}{H_{\mathrm{c2l}}}

\definecolor{racblue}{HTML}{1F77B4}
\definecolor{racgreen}{HTML}{2E8B57}
\definecolor{racorange}{HTML}{D97706}
\definecolor{racred}{HTML}{B23A48}
\definecolor{racgray}{HTML}{F2F4F7}
\definecolor{racmidgray}{HTML}{7A8793}
\definecolor{racdark}{HTML}{263238}
\definecolor{racgrid}{HTML}{D9DEE3}

\newcommand{\racplotfont}{\normalfont\fontsize{8}{9.6}\selectfont}
\newcommand{\racplottickfont}{\racplotfont}
\newcommand{\racplotlabelfont}{\racplotfont}
\newcommand{\racplottitlefont}{\racplotfont}
\newcommand{\racplotlegendfont}{\racplotfont}
\tikzset{
  rac model glm/.style={
    draw=racblue,
    solid
  },
  rac model qwen/.style={
    draw=racgreen,
    dash pattern=on 2.2pt off 1.1pt
  },
  rac model llama/.style={
    draw=racorange,
    dash pattern=on 0.35pt off 0.95pt
  }
}
\pgfplotsset{
  rac data plot/.style={
    axis line style={draw=racdark!82,line width=0.40pt},
    tick style={draw=racdark!82,line width=0.40pt},
    grid style={draw=racgrid,dashed,line width=0.30pt},
    tick label style={font=\racplottickfont,text=racdark},
    label style={font=\racplotlabelfont,text=racdark},
    title style={font=\racplottitlefont,text=racdark},
    legend style={
      draw=none,
      fill=none,
      font=\racplotlegendfont,
      text=racdark
    }
  },
  rac method raw/.style={
    draw=racmidgray!90!black,
    fill=racmidgray!38
  },
  rac method rac/.style={
    draw=racblue!90!black,
    fill=racblue!72
  },
  rac method topk eight/.style={
    draw=racorange!90!black,
    fill=racorange!62,
    postaction={
      pattern=north east lines,
      pattern color=racorange!82!black
    }
  },
  rac method topk four/.style={
    draw=racred!90!black,
    fill=racred!58,
    postaction={
      pattern=north west lines,
      pattern color=racred!82!black
    }
  }
}

\title{RAC: Reference-Aware Activation Compression\\
for Communication-Efficient Split LLM Inference}

\newif\ifshowauthors
\showauthorstrue
\ifshowauthors
\IEEEoverridecommandlockouts
\author{%
\IEEEauthorblockN{Guotao Yang, Mingxi Zhao, Haopeng Li, Zhengchao Wang, Sheng Chen, Yitao Hu\IEEEauthorrefmark{1}\thanks{\textsuperscript{*}Corresponding author: Yitao Hu (e-mail: \texttt{yitao@tju.edu.cn}).}, and Keqiu Li}
\IEEEauthorblockA{Tianjin University, China}}
\makeatletter
\def\@IEEEauthorblockconfadjspace{-1.25em}
\def\@IEEENORMtitlevspace{0.25\baselineskip}
\def\@IEEEMINtitlevspace{0.25\baselineskip}
\makeatother
\fi

\begin{document}
\bstctlcite{BSTcontrol}
\maketitle

\begin{abstract}
Large language model (LLM) agents repeatedly process long, privacy-sensitive contexts, while cloud-only deployment exposes user data beyond the trusted endpoint and fully local deployment often requires costly hardware. Split inference offers a middle ground by executing the model head, tail, and tools locally and the middle layers in the cloud, but its local--cloud--local path transfers boundary hidden states at every invocation and creates a critical communication bottleneck. We present \system, a reference-aware codec that retrieves exact-token historical spans for prefill uplinks, reuses the reconstructed uplink state for same-round prefill downlinks, and generates boundary-specific decode references with lightweight causal predictors. RAC applies grouped affine alignment and calibrated residual quantization with optional prefill outliers, while sender-side wire-format reconstruction synchronizes subsequent references and offline calibration accounts for quality and packed representation costs. Across three models and nine evaluated model--link pairs, Raw-to-RAC mean time to first token (TTFT) and time per output token (TPOT) ratios are 1.24--2.72$\times$ and 1.01--2.79$\times$, while the 12 non-perplexity task-score changes range from $-0.40$ to $+2.50$ points.
\end{abstract}

\begin{IEEEkeywords}
large language models, split inference, activation compression, residual quantization
\end{IEEEkeywords}

\section{Introduction}

Large language models (LLMs) are increasingly used as agents that call tools, consume returned data, and revise their plans, repeatedly extending the context with reasoning traces, documents, and tool outputs~\cite{brown2020gpt3,yao2023react,schick2023toolformer}. Current LLM services are predominantly deployed entirely in the cloud, benefiting from its abundant computing resources. However, this deployment requires users to upload substantial amounts of data to remote servers, raising privacy concerns. Although fully local LLM deployment keeps plaintext inputs on the trusted endpoint, the high cost of the required hardware often makes it impractical.

\begin{figure}[t]
    \centering
    \includegraphics[width=0.95\linewidth]{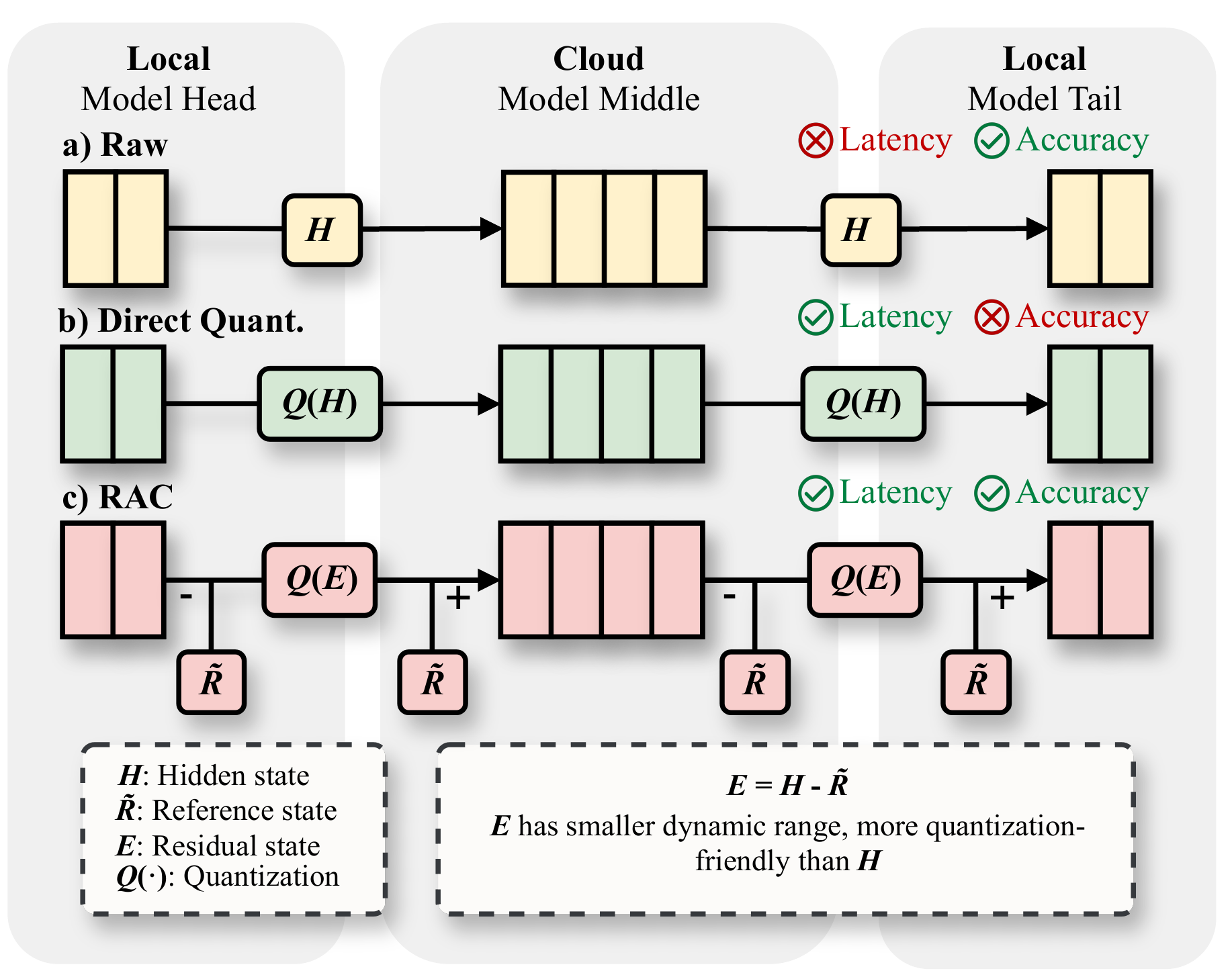}
    \caption{Comparison with baseline methods.}
    \label{fig:baseline_compare}
\end{figure}

The demand for both abundant cloud computing resources and strong privacy at the edge has given rise to edge--cloud split inference~\cite{kang2017neurosurgeon,eshatifar2021jointdnn}. In the three-way organization shown in Fig.~\ref{fig:baseline_compare}, a trusted local device executes the model head and tail while the cloud executes the middle layers. This arrangement retains plaintext inputs, logits, and tool invocations locally. Existing work strengthens the privacy of edge--cloud split inference by learning noise that suppresses information in transmitted features and by applying privacy-oriented pruning to retain task-relevant information while reducing sensitive content~\cite{mireshghallah2020shredder,ding2024patrol}.

This deployment introduces an \textbf{online communication bottleneck}. During prefill, each boundary carries a full-sequence hidden state, making time to first token (TTFT) sensitive to prompt length and available bandwidth. During decode, every generated token requires an uplink and a downlink for a one-token hidden state, making time per output token (TPOT) sensitive to round-trip time, serialization, and small-message overhead. Multi-round agent workloads amplify both costs. Direct quantization, sparsification, and mixed-precision compression can degrade inference accuracy. Direct quantization and the low-bit components of mixed-precision representations are sensitive to long-tailed channel distributions and outliers, while sparsification loses information by discarding activation values. Some sparsification and mixed-precision schemes additionally introduce indices, masks, and high-precision metadata~\cite{dettmers2022llmint8,xiao2023smoothquant,yao2022zeroquant}.

These limitations motivate transforming the hidden-state value distribution into a more concentrated distribution with fewer outliers before transmission so that the transformed state can be quantized more accurately. In Transformer-based LLM inference, \textbf{subtracting a similar reference hidden state from the current state can produce such a concentrated residual while allowing the receiver to reconstruct an approximation of the original state}. Practical agent workflows provide three sources of reference states: repeated system prompts, tool schemas, and stable context spans across rounds; the two boundary states linked by the Transformer blocks within one forward pass; and adjacent decode steps that share almost the entire causal context.

Turning these reuse opportunities into a practical codec raises \textbf{two key questions}. First, how can each phase and direction obtain a reference that is both inexpensive to construct and sufficiently similar to the current boundary state? Second, how can reference coding preserve task quality while keeping alignment, quantization, outlier handling, metadata, and critical-path processing costs low enough to retain its communication benefit?

We present \system, a reference-aware boundary-compression system for split LLM inference. RAC retrieves matching historical spans for prefill uplinks, reuses the shared reconstructed uplink state for same-round prefill downlinks, and predicts the current decode state from the preceding activation. The sender fits a grouped affine mapping from each reference to the current hidden state, quantizes the residual with a calibrated 4-bit or 8-bit representation and optional outlier payload, and locally reconstructs the state that the receiver obtains. Offline calibration chooses phase- and direction-specific configurations that meet the quality target while accounting for packed values, scales, alignment coefficients, and outlier metadata. RAC does not change model partitioning, the attention key-value cache, sampling, or tool semantics.

This paper makes the following contributions:
\begin{itemize}
  \item We develop a hidden-state codec that quantizes residuals relative to phase-specific reference states.
  \item We implement the codec for the two boundary transfers of three-way split inference, including phase-aware reference construction, alignment, residual quantization, and sender-side reconstruction.
  \item We evaluate \system on three LLMs using task quality, end-to-end latency, component ablations, activation-payload ratios, and stage-level runtime costs. Across nine model--link pairs, the Raw-to-RAC mean TTFT ratios are 1.24--2.72$\times$.
\end{itemize}

\section{Background}

\subsection{Large Language Model Inference}

Transformer-based LLMs process an input sequence through stacked attention and feed-forward blocks and then generate output tokens autoregressively. Inference has two phases~\cite{agrawal2024sarathi,zhong2024distserve}: prefill processes the complete prompt in parallel to produce the first output token, whereas decode appends one token at a time while reusing the attention key-value state of earlier tokens. Consequently, prefill operates on full-sequence hidden states and decode repeatedly operates on a single-token hidden state. These phase-dependent tensor shapes and execution patterns determine the computational and memory characteristics of LLM inference.

\subsection{Split LLM Inference Model}

\begin{figure*}[t]
\centering
\begingroup
\pgfplotsset{
  similarity glm/.style={
    color=racblue,solid,line width=0.90pt,
    mark=*,mark size=1.35pt
  },
  similarity qwen/.style={
    color=racgreen,solid,line width=0.90pt,
    mark=square*,mark size=1.25pt
  },
  similarity llama/.style={
    color=racorange,solid,line width=0.90pt,
    mark=triangle*,mark size=1.45pt
  }
}
\begin{tikzpicture}
\begin{groupplot}[
  rac data plot,
  group style={group size=4 by 1,horizontal sep=0.48cm},
  width=3.66cm,
  height=2.65cm,
  scale only axis,
  xmin=0.5,
  xmax=19.5,
  ymin=0,
  ymax=1,
  xtick={1,5,9,13,17},
  ytick={0,0.2,0.4,0.6,0.8,1},
  ymajorgrids,
  legend columns=3,
  legend style={
    /tikz/every even column/.append style={column sep=2.4mm}
  }
]
\nextgroupplot[legend to name=similaritylegend]
\addplot+[similarity glm] coordinates {
  (1,0.944578) (3,0.863694) (5,0.746769) (7,0.700109) (9,0.636483)
  (11,0.588569) (13,0.571184) (15,0.577269) (17,0.566784) (19,0.607685)
};
\addlegendentry{GLM}
\addplot+[similarity qwen] coordinates {
  (1,0.963488) (3,0.880497) (5,0.822167) (7,0.782766) (9,0.790028)
  (11,0.764471) (13,0.736923) (15,0.721047) (17,0.708600) (19,0.705591)
};
\addlegendentry{Qwen}
\addplot+[similarity llama] coordinates {
  (1,0.976023) (3,0.903992) (5,0.819410) (7,0.708773) (9,0.671038)
  (11,0.592012) (13,0.558522) (15,0.474196) (17,0.432876) (19,0.380585)
};
\addlegendentry{Llama}

\nextgroupplot[yticklabels=\empty]
\addplot+[similarity glm] coordinates {
  (1,0.068595) (3,0.162097) (5,0.210432) (7,0.269679) (9,0.332489)
  (11,0.376841) (13,0.477643) (15,0.615579) (17,0.756411) (19,0.907041)
};
\addplot+[similarity qwen] coordinates {
  (1,0.583012) (3,0.625332) (5,0.630642) (7,0.678021) (9,0.739570)
  (11,0.768476) (13,0.782341) (15,0.815354) (17,0.843387) (19,0.873694)
};
\addplot+[similarity llama] coordinates {
  (1,0.042051) (3,0.082247) (5,0.106793) (7,0.132440) (9,0.174785)
  (11,0.203365) (13,0.221108) (15,0.234484) (17,0.269585) (19,0.316758)
};

\nextgroupplot[yticklabels=\empty]
\addplot+[similarity glm] coordinates {
  (1,0.368706) (3,0.369831) (5,0.449754) (7,0.514736) (9,0.506446)
  (11,0.517136) (13,0.561161) (15,0.591342) (17,0.650681) (19,0.700857)
};
\addplot+[similarity qwen] coordinates {
  (1,0.624742) (3,0.641882) (5,0.646244) (7,0.667529) (9,0.665264)
  (11,0.690471) (13,0.691092) (15,0.677211) (17,0.684069) (19,0.698866)
};
\addplot+[similarity llama] coordinates {
  (1,0.381215) (3,0.445947) (5,0.492400) (7,0.499001) (9,0.539576)
  (11,0.512002) (13,0.506342) (15,0.493343) (17,0.465984) (19,0.466046)
};

\nextgroupplot[yticklabels=\empty]
\addplot+[similarity glm] coordinates {
  (1,0.658263) (3,0.570005) (5,0.590617) (7,0.610060) (9,0.621578)
  (11,0.541788) (13,0.583199) (15,0.615825) (17,0.692412) (19,0.669199)
};
\addplot+[similarity qwen] coordinates {
  (1,0.799134) (3,0.752490) (5,0.717762) (7,0.714092) (9,0.733716)
  (11,0.745896) (13,0.736989) (15,0.734931) (17,0.754913) (19,0.749191)
};
\addplot+[similarity llama] coordinates {
  (1,0.486423) (3,0.489421) (5,0.482087) (7,0.481044) (9,0.457245)
  (11,0.462374) (13,0.456751) (15,0.477943) (17,0.478530) (19,0.454194)
};
\end{groupplot}
\node at ($(group c2r1.north)!0.5!(group c3r1.north)+(0,0.58cm)$)
  {\ref{similaritylegend}};
\node[font=\racplottitlefont,anchor=north]
  at ($(group c1r1.south)+(0,-0.90cm)$) {(a) Across-input similarity};
\node[font=\racplottitlefont,anchor=north]
  at ($(group c2r1.south)+(0,-0.90cm)$) {(b) Same-round cross-layer};
\node[font=\racplottitlefont,anchor=north]
  at ($(group c3r1.south)+(0,-0.90cm)$) {(c) Adjacent-step uplink};
\node[font=\racplottitlefont,anchor=north]
  at ($(group c4r1.south)+(0,-0.90cm)$) {(d) Adjacent-step downlink};
\node[font=\racplotlabelfont,rotate=90]
  at ($(group c1r1.west)+(-0.76cm,0)$) {Cosine similarity};
\node[font=\racplotlabelfont]
  at ($(group c2r1.south)!0.5!(group c3r1.south)+(0,-0.58cm)$)
  {Number of head/tail layers};
\end{tikzpicture}
\endgroup
\caption{Cosine similarity between current boundary states and candidate references across models: repeated inputs, same-round cross-layer states, and adjacent decode states at the uplink and downlink boundaries.}
\label{fig:similarity-layer}
\end{figure*}
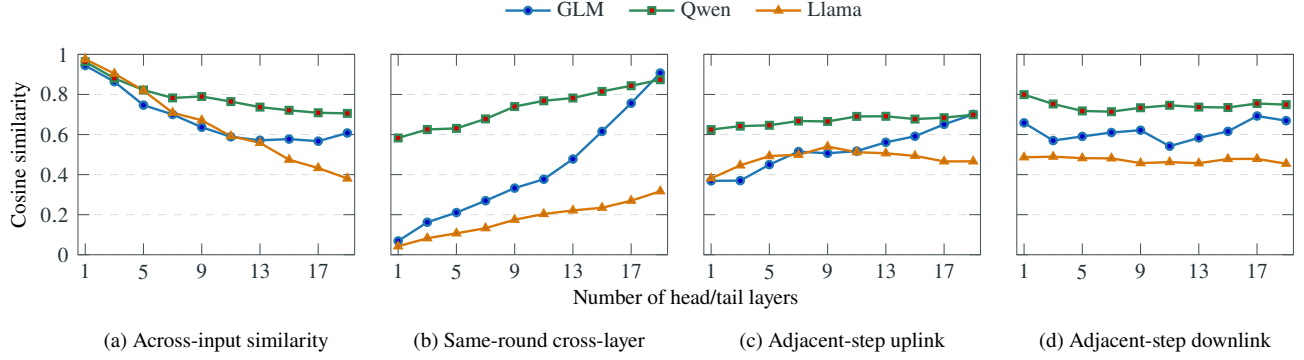

Split LLM inference partitions consecutive Transformer blocks across a local device and an edge or cloud server, which jointly execute the model by exchanging intermediate activations. It lets resource-constrained endpoints use remote computing and memory resources while retaining input and output functions locally, reducing local hardware requirements and limiting plaintext data exposure. This balance has motivated broad research in collaborative inference and distributed LLM serving~\cite{kang2017neurosurgeon,eshatifar2021jointdnn,borzunov2023petals,ye2025jupiter}. We consider a local--cloud--local organization in which the trusted endpoint executes the model head and tail while the cloud executes the middle layers. In this three-way organization, two cut points $l_1<l_2$ divide the LLM into contiguous functions
\begin{equation}
\begin{aligned}
f&=f_{\mathrm{tail}}\circ f_{\mathrm{middle}}\circ f_{\mathrm{head}},\\
\ltoC&=f_{\mathrm{head}}(X),\\
\ctoL&=f_{\mathrm{middle}}(\ltoC).
\end{aligned}
\label{eq:split-execution}
\end{equation}
Here, $X$ denotes the model input. The local device first executes the head layers and produces the local-to-cloud boundary state $\ltoC$. The cloud executes the middle layers on $\ltoC$ and returns the cloud-to-local state $\ctoL$. The local tail then continues the forward pass, produces logits, and samples the next token. Thus, $\ltoC$ and $\ctoL$ are the two intermediate tensors exchanged by this execution pattern.

The model revision, layer placement, and cut points determine the shapes and semantics of these boundary states. After either state crosses a cut, the receiving partition must obtain a shape-compatible tensor before executing the next unchanged model function. Boundary activations are distinct from model parameters, plaintext tokens, logits, tool data, and the persistent attention key-value cache.

\subsection{Boundary Communication Across Inference Phases}

LLM inference consists of prefill followed by autoregressive decode~\cite{agrawal2024sarathi,zhong2024distserve}. During prefill, one $[B,N,d]$ hidden state crosses each boundary: one is transferred on the uplink and the other on the downlink, where $B$, $N$, and $d$ denote batch size, input length, and hidden width. The communication volume therefore grows with the prompt and primarily affects TTFT. At each decode step, one $[B,1,d]$ state crosses each boundary. These two causally ordered boundary crossings recur for every generated token. Decode communication is consequently more sensitive to round-trip time, serialization, and small-message overhead~\cite{borzunov2023petals}.

\section{Motivation}

\subsection{Differencing Concentrates Data Distributions}

Existing activation-compression methods encode the current hidden state directly. Uniform low-bit quantization must cover the tensor's full dynamic range, so a few extreme values enlarge the quantization step for the remaining values; global quantization is especially sensitive because one scale covers the entire tensor. Sparsification avoids representing many values but discards information and may additionally transmit indices or masks, while mixed-precision methods retain selected outliers at the cost of high-precision values and metadata~\cite{dettmers2022llmint8,xiao2023smoothquant,yao2022zeroquant}. Directly reducing precision or retaining only selected values therefore creates a fundamental tension between communication cost and preservation of the information needed by downstream layers.

A complementary approach is to transform the coded distribution before quantization. Suppose that the sender and receiver construct the same aligned reference $\widetilde R$ for the current hidden state $H$. The sender encodes the residual $H-\widetilde R$, and the receiver reconstructs the state as
\begin{equation}
E=H-\widetilde R,\qquad \widehat H=\widetilde R+\widehat E,
\label{eq:motivation-residual}
\end{equation}
where $\widehat E$ is the dequantized residual. When $\widetilde R$ captures the dominant structure and extreme values of $H$, $E$ has a narrower, more concentrated distribution than $H$, so the same bit width can represent it with a finer step. The potential benefit therefore depends on whether each transfer can obtain an inexpensive shared reference that is sufficiently similar to $H$, and whether the resulting residual can be encoded without excessive error or auxiliary cost. These questions motivate the phase-specific opportunities examined next.

\subsection{Phase-Specific Reference Opportunities}

\textbf{Repeated spans across rounds.} Agent workloads often reuse system prompts, tool declarations, control tokens, and portions of the conversation history. Matching token spans therefore provide a low-cost lookup key for historical prefill-uplink states. In Fig.~\ref{fig:similarity-layer}(a), across-input similarity begins near 0.95 for all three models but decreases as the number of head or tail layers grows, ending near 0.70 for Qwen, 0.60 for GLM, and 0.38 for Llama. Repeated spans thus provide useful structure, but token equality does not imply identical activations because causal prefixes and token positions can differ even under a fixed model revision and cut configuration; the matched reference must therefore be aligned before residual formation.

\textbf{Same-round cross-boundary structure.} Fig.~\ref{fig:similarity-layer}(b) shows that Qwen's cross-layer similarity rises from roughly 0.58 to 0.88 and GLM's from roughly 0.06 to above 0.90 as fewer middle layers separate the boundaries; Llama rises more modestly from about 0.04 to 0.32. Because the downlink state $\ctoL$ is obtained by applying the cloud-side middle layers to the uplink state $\ltoC$, both endpoints already possess the reconstructed uplink state when the downlink is encoded. The model-dependent range shows that this same-round reference avoids historical lookup but still requires fitted grouped affine alignment rather than an identity mapping.

\textbf{Continuity across decode steps.} Consecutive-step similarity is consistently positive but differs by boundary. Fig.~\ref{fig:similarity-layer}(c) and (d) show uplink ranges of about 0.62--0.70 for Qwen, 0.37--0.70 for GLM, and 0.38--0.54 for Llama, compared with downlink ranges of roughly 0.71--0.80, 0.54--0.69, and 0.45--0.49, respectively. Consecutive steps use the same layer weights and contexts that differ only by the newest token, providing stable predictive structure for constructing boundary-specific references. Because activation structures vary across boundaries and models, separate predictors are required. The first decode step has no preceding activation and is therefore transmitted directly.

\subsection{Key Questions}

The observations above raise two key questions.\enspace\textbf{Q1: How can RAC construct effective phase-aware references without adding serial preparation?} Prefill uplinks, prefill downlinks, and decode use different reference sources. RAC must make each reference available before its boundary activation is ready.

\textbf{Q2: How can RAC align and encode residuals under quality and cost constraints?} A reference that appears similar need not quantize well: cosine similarity ignores magnitude and offset, and outliers can still enlarge the residual range. Alignment, finer groups, and explicit outlier values improve reconstruction but add computation and metadata. The system must therefore select phase-specific alignment and quantization configurations that preserve task quality while keeping the packed representation and exposed processing cost low. Section~\ref{sec:rac-design} addresses Q1 through overlapped reference preparation, and Q2 through quality- and cost-constrained aligned residual encoding.

\section{RAC Design}
\label{sec:rac-design}

\begin{figure*}[t]
\centering
\includegraphics[width=\textwidth]{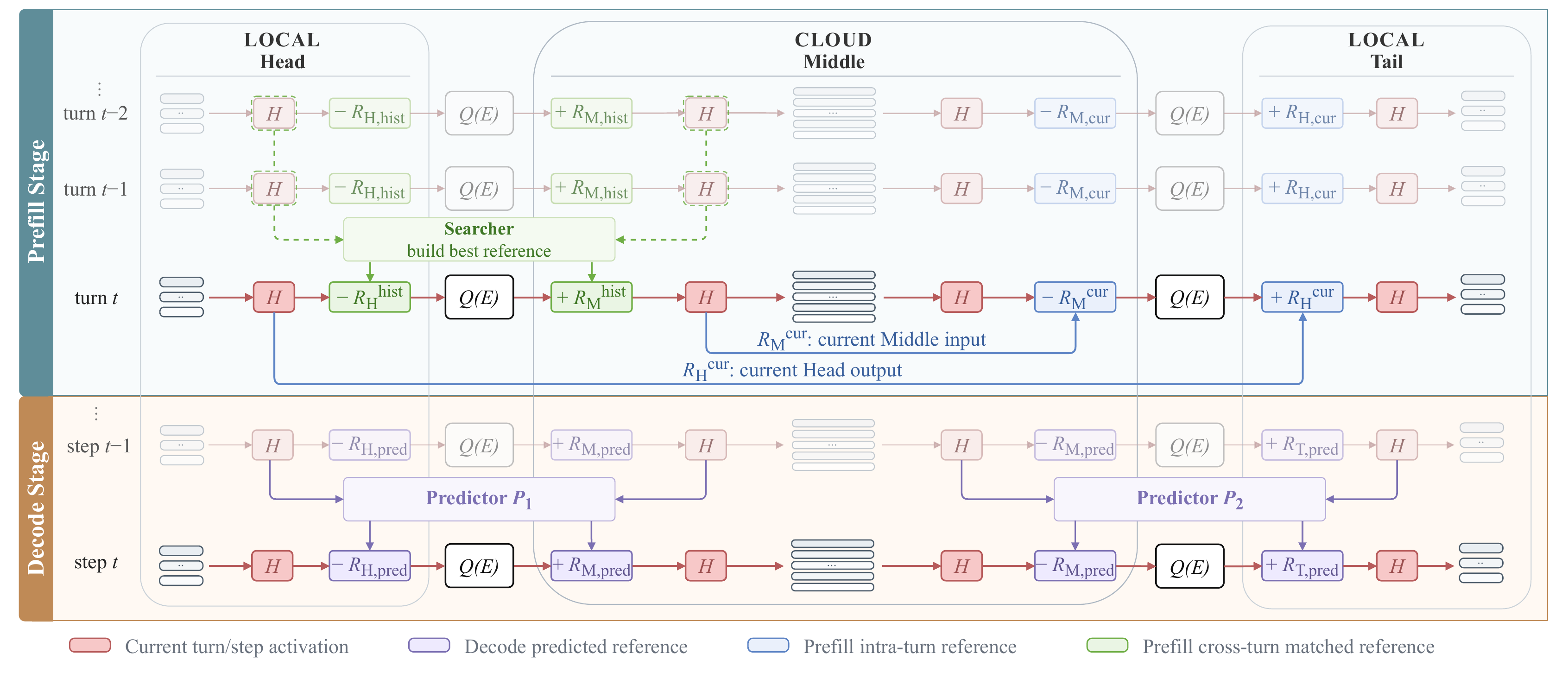}
\caption{RAC overview across prefill and decode. Historical-span search supplies the prefill uplink reference, the reconstructed uplink state supplies the same-round downlink reference, and boundary-specific predictors construct decode references from preceding activations. Each referenced transfer applies affine-aligned residual quantization.}
\label{fig:rac-overview}
\end{figure*}

\subsection{Overview}

As shown in Fig.~\ref{fig:rac-overview}, \system augments the local-head--cloud-middle--local-tail execution path with a prefill searcher, two decode predictors, and an aligned residual codec at each boundary. The searcher builds references from repeated historical spans for prefill uplinks, while the shared reconstructed uplink state serves as the reference for prefill downlinks. The predictors generate references for the two decode directions from their preceding boundary states. The codec aligns each reference, quantizes the residual, and reconstructs the transmitted activation using phase- and direction-specific parameters selected by offline quality and packed-cost calibration.

During prefill, the local head generates an uplink activation, and the searcher selects historical states for matched spans. Among uncovered positions, repeated tokens may reuse earlier reconstructed activations from the current sequence, while the remaining positions are encoded directly. The sender encodes residuals against the selected references, and the cloud reconstructs the activation, executes the middle layers, and returns a downlink encoded against the shared reconstructed uplink state. During decode, the first step is transferred directly because no preceding state exists; later steps use the two predictors to generate boundary-specific references and encode the corresponding residuals.

\subsection{Phase-Aware Reference Construction}

\textbf{Insight and approach.} Reference similarity is phase-, direction-, model-, and split-dependent. RAC retrieves repeated spans for prefill uplinks, reuses the reconstructed uplink for prefill downlinks, and predicts separate decode references. Historical-span selection and prefetch overlap local-head execution, the same-round downlink reference is available after uplink reconstruction, and decode prediction overlaps the remaining model computation or communication. These schedules make each reference available before residual formation.

\subsubsection{Repeated-Span Retrieval for Prefill Uplinks}

\begin{figure*}[t]
\centering
\begingroup
\pgfplotsset{
  predictor previous/.style={
    color=racmidgray,dashed,line width=0.90pt,
    mark=o,mark size=1.35pt,mark options={solid,fill=white}
  },
  predictor reference/.style={
    color=racblue,solid,line width=0.90pt,
    mark=triangle*,mark size=1.55pt
  }
}
\begin{tikzpicture}
\begin{groupplot}[
  rac data plot,
  group style={
    group size=3 by 2,
    horizontal sep=0.72cm,
    vertical sep=0.90cm
  },
  width=4.72cm,
  height=2.55cm,
  scale only axis,
  xmin=0.7,
  xmax=10.3,
  ymin=0,
  ymax=1,
  xtick={1,3,5,7,9},
  ytick={0,0.2,0.4,0.6,0.8,1},
  ymajorgrids,
  legend style={
    legend columns=2,
    /tikz/every even column/.append style={column sep=2.5mm}
  },
]

\nextgroupplot[
  xticklabels=\empty,
  legend to name=predictorlegend
]
\addplot+[predictor previous] coordinates {
  (1,0.482526) (2,0.443548) (3,0.411526) (4,0.368605) (5,0.387606)
  (6,0.503665) (7,0.435579) (8,0.409815) (9,0.539973) (10,0.388582)
};
\addlegendentry{Previous activation}
\addplot+[predictor reference] coordinates {
  (1,0.657827) (2,0.627067) (3,0.590334) (4,0.610063) (5,0.596901)
  (6,0.637206) (7,0.628741) (8,0.641447) (9,0.731844) (10,0.576186)
};
\addlegendentry{Predicted reference}

\nextgroupplot[
  xticklabels=\empty,
  yticklabels=\empty
]
\addplot+[predictor previous] coordinates {
  (1,0.585175) (2,0.656826) (3,0.621309) (4,0.646420) (5,0.657725)
  (6,0.618603) (7,0.619134) (8,0.632533) (9,0.620570) (10,0.664814)
};
\addplot+[predictor reference] coordinates {
  (1,0.757818) (2,0.780268) (3,0.773644) (4,0.835131) (5,0.807499)
  (6,0.765440) (7,0.791672) (8,0.779893) (9,0.765931) (10,0.825410)
};

\nextgroupplot[
  xticklabels=\empty,
  yticklabels=\empty
]
\addplot+[predictor previous] coordinates {
  (1,0.474360) (2,0.511321) (3,0.548656) (4,0.456420) (5,0.455108)
  (6,0.532689) (7,0.401876) (8,0.580536) (9,0.481600) (10,0.529642)
};
\addplot+[predictor reference] coordinates {
  (1,0.716702) (2,0.709228) (3,0.714180) (4,0.706379) (5,0.652389)
  (6,0.764337) (7,0.591222) (8,0.699434) (9,0.721217) (10,0.709920)
};

\nextgroupplot
\addplot+[predictor previous] coordinates {
  (1,0.643067) (2,0.682177) (3,0.510195) (4,0.641077) (5,0.427901)
  (6,0.735734) (7,0.667254) (8,0.559226) (9,0.627734) (10,0.615158)
};
\addplot+[predictor reference] coordinates {
  (1,0.695992) (2,0.742041) (3,0.584412) (4,0.665725) (5,0.501443)
  (6,0.781454) (7,0.702804) (8,0.663671) (9,0.686512) (10,0.674173)
};

\nextgroupplot[yticklabels=\empty]
\addplot+[predictor previous] coordinates {
  (1,0.701322) (2,0.732696) (3,0.652984) (4,0.740188) (5,0.701368)
  (6,0.768755) (7,0.725888) (8,0.628900) (9,0.598745) (10,0.757002)
};
\addplot+[predictor reference] coordinates {
  (1,0.831676) (2,0.828503) (3,0.784589) (4,0.875986) (5,0.833193)
  (6,0.851695) (7,0.873152) (8,0.770165) (9,0.691695) (10,0.850468)
};

\nextgroupplot[yticklabels=\empty]
\addplot+[predictor previous] coordinates {
  (1,0.443436) (2,0.564147) (3,0.559945) (4,0.519017) (5,0.476702)
  (6,0.572657) (7,0.408234) (8,0.565902) (9,0.516304) (10,0.535958)
};
\addplot+[predictor reference] coordinates {
  (1,0.615358) (2,0.608376) (3,0.635006) (4,0.660794) (5,0.625788)
  (6,0.689566) (7,0.485267) (8,0.651056) (9,0.615602) (10,0.637722)
};

\end{groupplot}
\node[font=\racplottitlefont,anchor=north]
  at ($(group c1r1.south)+(0,-0.16cm)$) {(a) GLM uplink};
\node[font=\racplottitlefont,anchor=north]
  at ($(group c2r1.south)+(0,-0.16cm)$) {(b) Qwen uplink};
\node[font=\racplottitlefont,anchor=north]
  at ($(group c3r1.south)+(0,-0.16cm)$) {(c) Llama uplink};
\node[font=\racplottitlefont,anchor=north]
  at ($(group c1r2.south)+(0,-0.94cm)$) {(d) GLM downlink};
\node[font=\racplottitlefont,anchor=north]
  at ($(group c2r2.south)+(0,-0.94cm)$) {(e) Qwen downlink};
\node[font=\racplottitlefont,anchor=north]
  at ($(group c3r2.south)+(0,-0.94cm)$) {(f) Llama downlink};
\node[font=\racplotlabelfont,rotate=90]
  at ($(group c1r1.west)!0.5!(group c1r2.west)+(-0.82cm,0)$)
  {Cosine similarity};
\node[font=\racplotlabelfont]
  at ($(group c2r2.south)+(0,-0.62cm)$)
  {Input index};
\node
  at ($(group c2r1.north)+(0,0.61cm)$)
  {\ref{predictorlegend}};
\end{tikzpicture}
\endgroup
\caption{Per-input cosine similarity between the current decode boundary state and either the immediately preceding activation or the predictor-generated reference. Results are shown for both split boundaries of GLM, Qwen, and Llama.}
\label{fig:predictor-cosine-similarity}
\end{figure*}
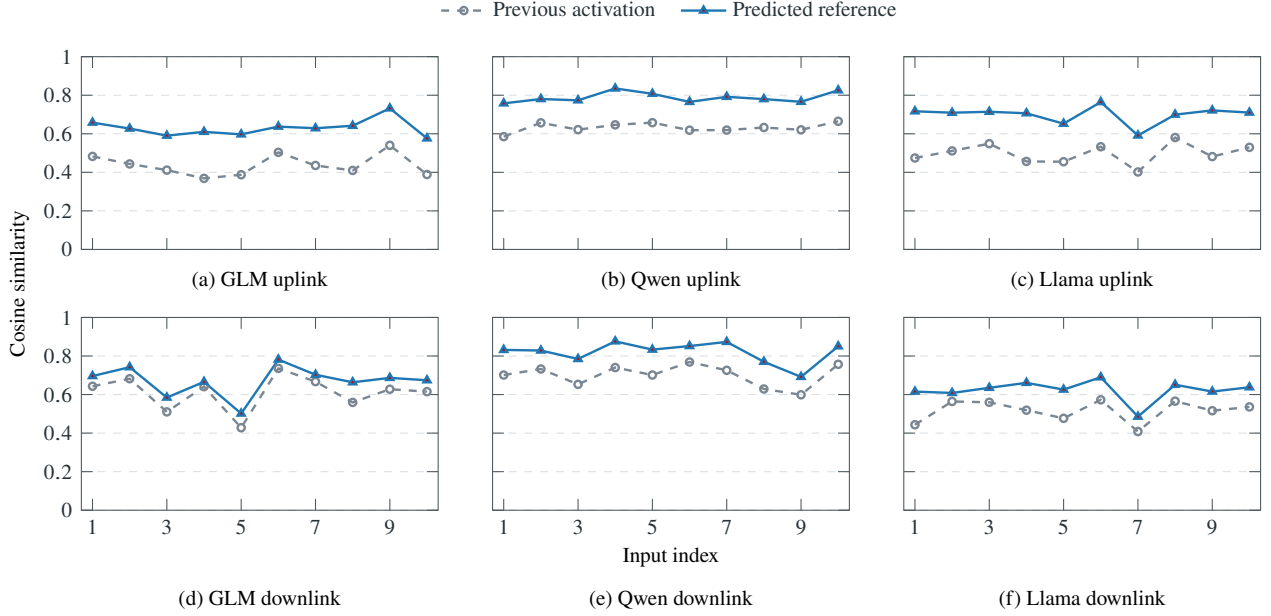

Let $x_{1:N}$ and $H$ denote the current prompt and prefill-uplink state. Rolling hashes index $w$-token windows in earlier requests and extend each hit to its longest exact match. At position $i$, candidate $c=(r,j,L)\in\mathcal C_i$ satisfies $x_{i:i+L-1}=x^{(r)}_{j:j+L-1}$ and $L\ge w$, where $r$, $j$, and $L$ identify the historical request, start position, and span length. RAC retains up to eight recent hits per hash bucket.

RAC performs deterministic left-to-right selection using token-match metadata:
\begin{equation}
c_i^\star=\operatorname*{lex\,arg\,max}_{c\in\mathcal C_i}
\bigl(L_c,\tau_c\bigr),
\label{eq:prefill-reference-selection}
\end{equation}
where $\tau_c$ denotes recency. RAC prioritizes the longest exact span and breaks ties by recency. After selecting $c_i^\star$, the scan advances to position $i+L_{c_i^\star}$. If $\mathcal C_i$ is empty, the scan advances to $i+1$. Positions covered by a selected span use the corresponding historical activation states as references. Among the uncovered positions, repeated tokens may reuse an earlier reconstructed activation from the current sequence as an intra-sequence reference; all remaining positions are encoded directly. Because selection depends only on tokens and historical metadata, the selected activation spans are prefetched while the local head executes. After $H$ becomes available, RAC aligns the selected references and forms the residuals.

\subsubsection{Same-Round References for Prefill Downlinks}

For the downlink of the same split execution, both endpoints already possess the reconstructed uplink state:
\begin{equation}
R_{\mathrm{cross}}=\widehat H_{\mathrm{l2c}}.
\label{eq:cross-reference}
\end{equation}
The cloud uses $R_{\mathrm{cross}}$ for $H_{\mathrm{c2l}}$ without historical lookup. Because the two tensors come from different layers and their similarity is model- and split-dependent, RAC does not assume element-wise closeness; grouped affine alignment transforms the shared state before residual formation.

\subsubsection{Predicted References for Decode}

The first decode step has no preceding activation and is therefore transmitted directly. From the second step onward, each boundary uses a dedicated one-layer causal Transformer trained on consecutive activations from the corresponding target layer. Training records target-layer activations and optimizes next-activation prediction within a bounded history; at inference, the predictor retains this bounded causal state and uses a key-value cache. For $t\ge 2$, the predicted reference at boundary $b$ is
\begin{equation}
R_{\mathrm{pred}}^{t,b}=P_{\phi_b}\!\left(\widehat H^{t-1,b};S^{t-1,b}\right),\qquad t\ge 2,
\label{eq:decode-predictor}
\end{equation}
where $S^{t-1,b}$ denotes the predictor's retained causal state. Separate parameters $\phi_b$ account for the observed difference between uplink and downlink similarity patterns.

Fig.~\ref{fig:predictor-cosine-similarity} validates this construction on ten evaluated inputs. Prediction improves cosine similarity for every model, input, and boundary; relative to direct reuse of the preceding activation, the mean gain is 0.156--0.201 for the uplink and 0.059--0.118 for the downlink, so the aggregate gain is not driven by a few favorable inputs. Predictors $P_1$ and $P_2$ serve the uplink and downlink boundaries, respectively, and their outputs are used only as communication references, never as model inputs or token-generation states.

\subsubsection{Reference Construction Scheduling}

Token hashing, exact-span selection, historical lookup, and selected-reference prefetch overlap local-head execution. After $H$ becomes available, alignment and residual formation, residual quantization, and sender-side reconstruction remain on the prefill-uplink sender path. Decode prediction overlaps the remaining model computation or communication after the preceding state becomes available. Section~\ref{sec:runtime-overhead} reports these exposed codec stages.

\subsection{Quality- and Cost-Constrained Aligned Residual Encoding}
\label{subsec:residual-encoding}

\textbf{Insight and approach.} A similar reference can preserve the activation's dominant structure, but cosine similarity alone does not correct channel-wise scale and offset or control residual outliers. RAC first fits grouped affine mappings, then quantizes the aligned residual with calibrated group sizes and optional prefill outliers. Online, the sender applies the same wire-format reconstruction as the receiver to keep subsequent references synchronized. Offline calibration reuses this reconstruction path to select phase- and direction-specific parameters that meet the quality target while accounting for value, scale, coefficient, and outlier bytes.

\subsubsection{Grouped Affine Alignment}
\label{subsubsec:alignment}

Assuming $g_a\mid d$, \system partitions the hidden dimension into $J_a=d/g_a$ alignment groups. For the $g_a$-channel slices $H_j$ and $R_j$ over the fitting positions, group $j$ solves
\begin{equation}
(\alpha_j^\star,\beta_j^\star)
=\operatorname*{arg\,min}_{\alpha,\beta}\,
\bigl\|H_j-(\alpha R_j+\beta)\bigr\|_{\mathrm F}^2.
\label{eq:affine-fit}
\end{equation}
For zero-variance $R_j$, $\alpha_j^\star=0$ and $\beta_j^\star=\operatorname{mean}(H_j)$. Both endpoints use the serialized coefficients $(\alpha_j,\beta_j)=\operatorname{Round}_{\mathrm{wire}}(\alpha_j^\star,\beta_j^\star)$, where $\operatorname{Round}_{\mathrm{wire}}$ denotes rounding to the configured coefficient format:
\begin{subequations}
\label{eq:aligned-residual}
\begin{align}
\widetilde R_j&=\alpha_jR_j+\beta_j,
\label{eq:aligned-reference}\\
E&=H-\widetilde R.
\label{eq:residual-definition}
\end{align}
\end{subequations}
Smaller $g_a$ improves fit but serializes more coefficient pairs; decode may therefore require coarser groups than prefill. Calibration selects $g_a$ per model, split, phase, and direction.

\subsubsection{Grouped Residual Quantization}

\system partitions $E$ into $g_q$-element groups. An optional mask $M_j\in\{0,1\}^{g_q}$ separates sparse outliers from the dense base:
\begin{equation*}
O_j=M_j\odot E_j,\qquad E_j^d=E_j-O_j.
\end{equation*}
Without outliers, $M_j=\mathbf{0}$. For $q\in\{4,8\}$, define
\begin{subequations}
\label{eq:residual-quantization}
\begin{align}
Q_{\max}&=2^{q-1}-1,\\
s_j^{\star,E}
&=\frac{\max\lvert E_j^d\rvert}{Q_{\max}},\qquad
\bar{s}_j^E=\operatorname{RoundUp}_{\mathrm{wire}}(s_j^{\star,E})
\ge s_j^{\star,E},\\
q_j&=\operatorname{clip}_{[-Q_{\max},Q_{\max}]}
\!\left(\lfloor E_j^d/\bar{s}_j^E\rceil\right).
\end{align}
\end{subequations}
All-zero groups carry a zero marker and skip division. Here, $\operatorname{RoundUp}_{\mathrm{wire}}$ rounds upward in the configured scale format, preventing dense-value clipping. Four-bit integer (INT4) quantization packs two values per byte, and each group carries its serialized scale. Calibration selects $(g_a,g_q,q,k_g)$ in outlier mode and $(g_a,g_q,q)$ otherwise, balancing local adaptation against scale, alignment, and outlier metadata, especially for single-token decode.

\subsubsection{Outlier Metadata and Calibration}

In outlier mode, $M_j$ selects the $k_g$ largest magnitudes per fixed group; the packet carries in-group offsets and $\widehat O_j=\operatorname{Round}_{\mathrm{FP16}}(O_j)$, where FP16 denotes the 16-bit floating-point format. Fixed group order avoids full coordinates. Calibration considers group size, outlier count, selection time, metadata bytes, and reconstruction quality.

For decode, \system disables mixed-precision outlier handling: it neither extracts outliers nor transmits selected values at their original precision, so the packet carries no outlier mask, offsets, or high-precision payload. The quantization group spans one complete token vector, i.e., $g_q=d$ for a $[B,1,d]$ boundary state. Consequently, each token requires only one quantization scale per batch element, minimizing the auxiliary data used by decode quantization.

\subsubsection{Reconstruction and Error}

The receiver resolves $R$, applies the serialized alignment, dequantizes the dense base, and restores outliers:
\begin{subequations}
\label{eq:residual-reconstruction}
\begin{align}
\widehat E_j
&=\operatorname{DQ}(q_j;\bar{s}_j^E)+\widehat O_j,
\label{eq:dense-residual-reconstruction}\\
\widehat H
&=\widetilde R+\widehat E.
\label{eq:boundary-reconstruction}
\end{align}
\end{subequations}
Here, $\operatorname{DQ}$ denotes dequantization using the serialized residual scale.
For a common reference $R$ and common serialized alignment coefficients,
\begin{equation}
H-\widehat H=E-\widehat E.
\label{eq:error-equivalence}
\end{equation}
Thus the reference changes the coded distribution but adds no independent error term.

For round-to-nearest, let $n_j^d$ count dense values. Eq.~\eqref{eq:residual-quantization} prevents dense clipping, and disjoint dense/outlier positions give
\begin{equation}
\begin{aligned}
\|E-\widehat E\|_{\mathrm F}^2
&=\sum_j\bigl\|E_j^d-\operatorname{DQ}(q_j;\bar{s}_j^E)\bigr\|_{\mathrm F}^2
+\sum_j\|O_j-\widehat O_j\|_{\mathrm F}^2\\
&\le\sum_j n_j^d\frac{(\bar{s}_j^E)^2}{4}
+\sum_j\|O_j-\widehat O_j\|_{\mathrm F}^2.
\end{aligned}
\label{eq:error-bound}
\end{equation}
For the same grouping, outlier-selection rule and count, and bit width, let $H_j^d$ be the dense base obtained by applying that rule to direct-coded $H_j$ (thus $H_j^d=H_j$ without outliers), and define $M_{X,j}^d=\max\lvert X_j^d\rvert$ for $X\in\{E,H\}$. If $M_{H,j}^d>0$, let
$\rho_j=M_{E,j}^d/M_{H,j}^d$; then before wire-scale rounding
\begin{equation}
\frac{n_j^d}{4}\!\left(\frac{M_{E,j}^d}{Q_{\max}}\right)^2
=\rho_j^2\frac{n_j^d}{4}\!\left(\frac{M_{H,j}^d}{Q_{\max}}\right)^2.
\label{eq:residual-range-gain}
\end{equation}
A uniform $0\le\rho_j\le\rho<1$ bounds each residual pre-rounding dense term by $\rho^2$ times its direct-coded counterpart; it does not guarantee a $\rho^2$ reduction of Eq.~\eqref{eq:error-bound}, because wire-scale rounding and FP16 outlier error remain. If $M_{H,j}^d=0$, the ratio is undefined and no range-reduction claim is made.

\section{Evaluation}
\label{sec:evaluation}

\subsection{Experiment Setup}
\label{subsec:experiment-setup}

\begin{table*}[!t]
\centering
\caption{End-to-end quality and activation-value payload (metadata/control excluded), normalized to Raw as prefill/decode. Accuracy, exact-match, and MathVerify values use a 0--100 percentage scale. Each derived row reports RAC minus Raw; an em dash denotes N/A. Arrows show the preferred direction; boldface and underlining mark the best and second-best quality values.}
\label{tab:precision-complete}
\begingroup
\footnotesize
\sisetup{
  mode=math,
  group-digits=none,
  retain-explicit-plus=true,
  detect-weight=true,
  table-number-alignment=center,
  table-text-alignment=center
}
\setlength{\tabcolsep}{1.7pt}
\renewcommand{\arraystretch}{1.00}
\begin{tabular}{@{}
  l
  l
  c
  @{\hspace{6pt}}
  *{5}{c}
  @{}}
\toprule
\multirow{2}{*}{\textbf{Model}} &
\multirow{2}{*}{\textbf{Method}} &
  \multirow{2}{*}{\shortstack{\textbf{Payload / Raw (\%)}\\Prefill / Decode}} &
  \mbox{\textbf{Language modeling}} &
  {\textbf{Commonsense}} &
  \multicolumn{3}{c}{\textbf{Mathematical reasoning}} \\
\cmidrule(lr){4-4}\cmidrule(lr){5-5}\cmidrule(lr){6-8}
& & &
  {\shortstack{\textbf{WikiText-2}\\PPL $\downarrow$}} &
  {\shortstack{\textbf{HellaSwag}\\Norm.\ Acc.\ $\uparrow$}} &
  {\shortstack{\textbf{GSM8K}\\Flexible EM $\uparrow$}} &
  {\shortstack{\textbf{GSM8K}\\Strict EM $\uparrow$}} &
  {\shortstack{\textbf{MATH-500}\\MathVerify $\uparrow$}} \\
\midrule
\multirow{7}{*}{\shortstack[l]{\textbf{GLM-4}\\9B-0414}} & Raw
  & 100 / 100
  & 40.77 & {\textbf{80.16}} &
    {\underline{80.59}} &
    {\underline{80.44}} &
    {\underline{56.00}} \\
& TopK (4-bit ratio)
  & 25 / 25
  & 123.00 & 50.73 & 14.63 & 12.05 & 5.00 \\
& TopK (8-bit ratio)
  & 50 / 50
  & 42.93 & 72.78 & 64.22 & 69.90 & 38.20 \\
& Global INT8
  & 50 / 50
  & {\textbf{31.87}} & 79.50 &
    {\textbf{80.89}} &
    {\textbf{80.82}} & 53.80 \\
& Global INT4
  & 25 / 25
  & 214.87 & 45.69 & 1.82 & 0.61 & 0.80 \\
\cmidrule(l){2-8}
& \textbf{\system (Ours)}
  & 27.8 / 25.0
  & {\underline{39.66}} &
    {\underline{80.10}} &
    80.21 & 80.36 & {\textbf{56.60}} \\
& \emph{$\Delta$ vs.\ Raw}
  & \textemdash
  & -1.11 & -0.06 & -0.38 & -0.08 & +0.60 \\

\midrule
\multirow{7}{*}{\shortstack[l]{\textbf{Qwen3}\\30B-A3B}} & Raw
  & 100 / 100
  & {\textbf{19.18}} &
    {\textbf{77.74}} &
    {\underline{86.05}} &
    {\underline{89.38}} &
    {\textbf{80.00}} \\
& TopK (4-bit ratio)
  & 25 / 25
  & 56.73 & 60.02 & 16.00 & 14.78 & 24.40 \\
& TopK (8-bit ratio)
  & 50 / 50
  & 26.80 & 72.27 & 75.13 & 74.68 & 68.80 \\
& Global INT8
  & 50 / 50
  & 20.62 & 76.63 & 84.38 & 86.96 & 77.60 \\
& Global INT4
  & 25 / 25
  & 1110.52 & 32.13 & 0.68 & 0.08 & 0.60 \\
\cmidrule(l){2-8}
& \textbf{\system (Ours)}
  & 27.8 / 25.0
  & {\underline{19.45}} &
    {\underline{77.42}} &
    {\textbf{87.11}} &
    {\textbf{89.76}} &
    {\underline{79.80}} \\
& \emph{$\Delta$ vs.\ Raw}
  & \textemdash
  & +0.27 & -0.32 & +1.06 & +0.38 & -0.20 \\

\midrule
\multirow{7}{*}{\shortstack[l]{\textbf{Llama-3.3}\\70B-Instruct}} & Raw
  & 100 / 100
  & {\textbf{13.46}} &
    {\underline{84.85}} &
    {\underline{91.89}} & 73.77 &
    {\textbf{65.80}} \\
& TopK (4-bit ratio)
  & 25 / 25
  & 36.17 & 80.83 & 81.43 & 42.23 & 53.40 \\
& TopK (8-bit ratio)
  & 50 / 50
  & 20.02 & 83.66 & 91.05 & 49.81 & 60.40 \\
& Global INT8
  & 50 / 50
  & 20.36 & 82.91 & 80.44 &
    {\textbf{85.22}} & 48.00 \\
& Global INT4
  & 25 / 25
  & 16932.98 & 26.35 & 1.90 & 0.00 & 0.80 \\
\cmidrule(l){2-8}
& \textbf{\system (Ours)}
  & 27.8 / 25.0
  & {\underline{13.54}} &
    {\textbf{84.87}} &
    {\textbf{92.04}} &
    {\underline{76.27}} &
    {\underline{65.40}} \\
& \emph{$\Delta$ vs.\ Raw}
  & \textemdash
  & +0.08 & +0.02 & +0.15 & +2.50 & -0.40 \\
\bottomrule
\end{tabular}
\endgroup
\end{table*}

\textbf{Testbed.}
Our testbed comprises a local platform equipped with two NVIDIA RTX 3090 GPUs and a cloud platform equipped with either one or four NVIDIA A800 GPUs. We use Linux Traffic Control to emulate heterogeneous wide-area network conditions~\cite{linux-tc}. Unless otherwise stated, the end-to-end latency experiments use a request concurrency of eight.

\textbf{Models.}
We evaluate RAC using three representative models: GLM-4-9B-0414, Qwen3-30B-A3B, and Llama-3.3-70B-Instruct~\cite{glm4-9b-0414-card,qwen3-report,llama33-card}. They span parameter scales from 9B to 70B and include both dense and mixture-of-experts architectures, allowing us to examine RAC across heterogeneous computation and memory characteristics under split inference. For compact labels in figures and discussion, GLM, Qwen, and Llama denote GLM-4-9B-0414, Qwen3-30B-A3B, and Llama-3.3-70B-Instruct, respectively.

\textbf{Datasets.}
For quality evaluation, we use WikiText-2, HellaSwag, GSM8K, and MATH-500~\cite{merity2016wikitext,zellers2019hellaswag,cobbe2021gsm8k,lightman2023verify}. WikiText-2 evaluates language-modeling quality, HellaSwag evaluates commonsense reasoning, and GSM8K and MATH-500 evaluate mathematical reasoning at different difficulty levels. For performance evaluation, we use ShareGPT~\cite{sharegpt}, a collection of publicly shared multi-turn conversations; our workload includes heterogeneous input and output lengths.

\textbf{Baselines.}
We compare RAC with five baselines: Raw, TopK (8-bit ratio), TopK (4-bit ratio), Global INT8 (global 8-bit integer quantization), and Global INT4 (global 4-bit integer quantization). The end-to-end quality evaluation includes all five baselines; the latency evaluation includes Raw, RAC, and the two TopK configurations; and the bandwidth-sensitivity evaluation compares RAC with Raw.

\textbf{Metrics.}
For performance, we report time to first token (TTFT), time per output token (TPOT), and output-token throughput. TTFT and TPOT are summarized using the mean, median, and 99th percentile. We also report the transferred activation-value payload relative to Raw, separately for prefill and decode; this payload ratio excludes metadata and control traffic. For quality, we report word-level perplexity (PPL) on WikiText-2, normalized accuracy on HellaSwag, flexible and strict exact match (EM) on GSM8K, and MathVerify accuracy on MATH-500. Accuracy and EM values use a 0--100 percentage scale. Lower latency and perplexity are better, whereas higher throughput and accuracy are better.

\subsection{End-to-End Quality}
\label{sec:end-to-end-quality}

Table~\ref{tab:precision-complete} compares end-to-end quality across the three evaluated models. RAC has the smallest absolute deviation from Raw among the five compressed methods in 14 of the 15 model--metric cells; the exception is GLM's GSM8K flexible exact-match accuracy, where global 8-bit integer quantization differs by 0.30 points and RAC by 0.38 points. RAC's non-perplexity changes range from $-0.40$ to $+2.50$ points, with the largest change occurring on Llama's GSM8K strict exact-match accuracy. Its relative WikiText-2 perplexity changes are $-2.72\%$, $+1.41\%$, and $+0.59\%$ for GLM, Qwen, and Llama, respectively; the direction of the change is not uniform across the three evaluated models.

The payload ratios in Table~\ref{tab:precision-complete} are constant across the evaluated 1k, 2k, and 8k sequence lengths, so we report normalized ratios rather than repeating the absolute megabyte counts. For all three models, RAC transfers approximately $27.8\%$ of the Raw prefill payload and $25.0\%$ of the Raw decode payload, corresponding to payload reductions of approximately $72.2\%$ and $75.0\%$, respectively. The 2.8-percentage-point prefill premium reflects its optional high-precision outliers, whereas decode disables outlier transmission and remains at the dense INT4 activation-value ratio.

The remaining methods are not quality-matched to Raw or RAC. Both TopK configurations reduce every higher-is-better score, with larger losses for TopK (4-bit ratio). Global INT8 is closer but inconsistent across models and metrics, while Global INT4 yields perplexity values from 214.87 to 16932.98 and near-zero GSM8K or MATH-500 scores in several cells. Accordingly, the TopK latency results represent different quality operating points rather than equal-quality comparisons.

\subsection{End-to-End Latency}

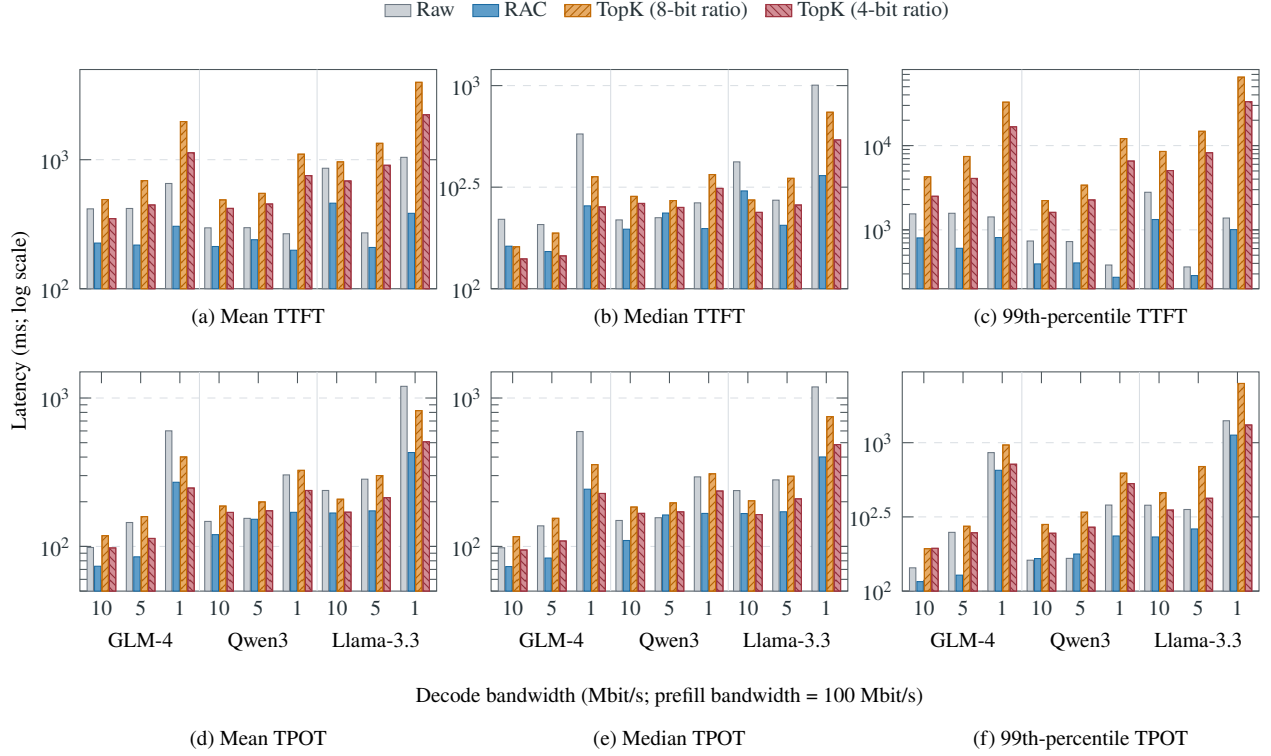
\begin{figure*}[t]
\centering
\begingroup
\pgfplotsset{
  latency raw/.style={
    ybar,bar width=2.4pt,bar shift=-4.2pt,mark=none,
    rac method raw,line width=0.40pt
  },
  latency rac/.style={
    ybar,bar width=2.4pt,bar shift=-1.4pt,mark=none,
    rac method rac,line width=0.40pt
  },
  latency topk eight/.style={
    ybar,bar width=2.4pt,bar shift=1.4pt,mark=none,
    rac method topk eight,line width=0.40pt
  },
  latency topk four/.style={
    ybar,bar width=2.4pt,bar shift=4.2pt,mark=none,
    rac method topk four,line width=0.40pt
  }
}
\newcommand{\latencymodellabels}{%
  \node[font=\racplottitlefont,anchor=north]
    at (axis description cs:0.17,-0.145) {GLM-4};
  \node[font=\racplottitlefont,anchor=north]
    at (axis description cs:0.50,-0.145) {Qwen3};
  \node[font=\racplottitlefont,anchor=north]
    at (axis description cs:0.83,-0.145) {Llama-3.3};
}
\begin{tikzpicture}
\begin{groupplot}[
  rac data plot,
  group style={
    group size=3 by 2,
    horizontal sep=0.72cm,
    vertical sep=1.10cm
  },
  width=4.72cm,
  height=2.90cm,
  scale only axis,
  ymode=log,
  log origin=infty,
  ymajorgrids,
  xmin=-0.55,
  xmax=8.55,
  xtick={0,1,2,3,4,5,6,7,8},
  xticklabels={10,5,1,10,5,1,10,5,1},
  clip=false,
]

\nextgroupplot[
  ymin=100,
  ymax=5000,
  xtick=\empty,
  legend to name=latencybarlegend,
  legend columns=4,
  legend style={
    legend image code/.code={
      \draw[#1] (0cm,-0.08cm) rectangle (0.34cm,0.08cm);
    },
    /tikz/every even column/.append style={column sep=1.6mm}
  }
]
\addplot+[latency raw] coordinates {
  (0,416.03) (1,418.85) (2,654.21)
  (3,296.49) (4,297.29) (5,266.8)
  (6,858.5) (7,271.02) (8,1043.16)
};
\addlegendentry{Raw}
\addplot+[latency rac] coordinates {
  (0,225.94) (1,218.02) (2,304.83)
  (3,212.49) (4,239.85) (5,198.83)
  (6,460.35) (7,208.87) (8,384.15)
};
\addlegendentry{RAC}
\addplot+[latency topk eight] coordinates {
  (0,491.35) (1,688.56) (2,1976.75)
  (3,488.43) (4,548.39) (5,1103.84)
  (6,962.53) (7,1339.88) (8,3980.61)
};
\addlegendentry{TopK (8-bit ratio)}
\addplot+[latency topk four] coordinates {
  (0,349.58) (1,446.03) (2,1131.61)
  (3,419.47) (4,453.76) (5,753.53)
  (6,684.47) (7,905.53) (8,2230.32)
};
\addlegendentry{TopK (4-bit ratio)}
\draw[racgrid,line width=0.4pt] (axis cs:2.5,100) -- (axis cs:2.5,5000);
\draw[racgrid,line width=0.4pt] (axis cs:5.5,100) -- (axis cs:5.5,5000);
\nextgroupplot[
  ymin=100,
  ymax=1200,
  xtick=\empty
]
\addplot+[latency raw] coordinates {
  (0,219.67) (1,207.04) (2,577.99)
  (3,218.19) (4,223.7) (5,264.7)
  (6,421.24) (7,272.82) (8,1005.1)
};
\addplot+[latency rac] coordinates {
  (0,161.97) (1,152.48) (2,255.87)
  (3,196.4) (4,235.86) (5,197.79)
  (6,303.1) (7,205.26) (8,360.61)
};
\addplot+[latency topk eight] coordinates {
  (0,160.92) (1,187.89) (2,356.12)
  (3,284.89) (4,271.77) (5,365)
  (6,273.48) (7,349.61) (8,739.65)
};
\addplot+[latency topk four] coordinates {
  (0,140.25) (1,145.15) (2,252.89)
  (3,263.02) (4,251.36) (5,311.87)
  (6,237.65) (7,258.35) (8,539.95)
};
\draw[racgrid,line width=0.4pt] (axis cs:2.5,100) -- (axis cs:2.5,1200);
\draw[racgrid,line width=0.4pt] (axis cs:5.5,100) -- (axis cs:5.5,1200);
\nextgroupplot[
  ymin=200,
  ymax=80000,
  xtick=\empty
]
\addplot+[latency raw] coordinates {
  (0,1547.89) (1,1572.77) (2,1423.79)
  (3,737.18) (4,726.18) (5,382.61)
  (6,2793.06) (7,362.97) (8,1381.18)
};
\addplot+[latency rac] coordinates {
  (0,802.26) (1,603.82) (2,810.22)
  (3,394.25) (4,405.00) (5,273.6)
  (6,1326.32) (7,286.76) (8,1006.91)
};
\addplot+[latency topk eight] coordinates {
  (0,4261.72) (1,7428.99) (2,32788.21)
  (3,2227.74) (4,3391.65) (5,12131.19)
  (6,8522.97) (7,14797.07) (8,65008.94)
};
\addplot+[latency topk four] coordinates {
  (0,2504) (1,4075.94) (2,16716.62)
  (3,1613.03) (4,2266.74) (5,6561.63)
  (6,5055.06) (7,8223.36) (8,33222.17)
};
\draw[racgrid,line width=0.4pt] (axis cs:2.5,200) -- (axis cs:2.5,80000);
\draw[racgrid,line width=0.4pt] (axis cs:5.5,200) -- (axis cs:5.5,80000);
\nextgroupplot[
  ymin=50,
  ymax=1500
]
\addplot+[latency raw] coordinates {
  (0,98.36) (1,144.93) (2,600.89)
  (3,147.42) (4,154.56) (5,303.33)
  (6,238.24) (7,283.84) (8,1199.04)
};
\addplot+[latency rac] coordinates {
  (0,73.53) (1,85.1) (2,270.62)
  (3,120.07) (4,152.38) (5,169.76)
  (6,167.99) (7,173.62) (8,429.25)
};
\addplot+[latency topk eight] coordinates {
  (0,118.14) (1,158.81) (2,400.94)
  (3,187.13) (4,199.55) (5,325.6)
  (6,208) (7,300.38) (8,822.12)
};
\addplot+[latency topk four] coordinates {
  (0,97.7) (1,113.2) (2,247.53)
  (3,169.5) (4,173.92) (5,238.23)
  (6,170.12) (7,213.14) (8,507.11)
};
\draw[racgrid,line width=0.4pt] (axis cs:2.5,50) -- (axis cs:2.5,1500);
\draw[racgrid,line width=0.4pt] (axis cs:5.5,50) -- (axis cs:5.5,1500);
\latencymodellabels

\nextgroupplot[
  ymin=50,
  ymax=1500
]
\addplot+[latency raw] coordinates {
  (0,97.95) (1,137.62) (2,594.27)
  (3,149.58) (4,156.11) (5,294.21)
  (6,237.84) (7,280.83) (8,1187.94)
};
\addplot+[latency rac] coordinates {
  (0,73.17) (1,83.51) (2,242.96)
  (3,109.63) (4,162.92) (5,166.91)
  (6,166.59) (7,171.21) (8,400.74)
};
\addplot+[latency topk eight] coordinates {
  (0,116.31) (1,154.66) (2,355.92)
  (3,184.31) (4,196.64) (5,308.63)
  (6,202.95) (7,297.24) (8,749.69)
};
\addplot+[latency topk four] coordinates {
  (0,94.6) (1,108.81) (2,227.46)
  (3,167.01) (4,170.99) (5,236.31)
  (6,163.98) (7,209.19) (8,485.67)
};
\draw[racgrid,line width=0.4pt] (axis cs:2.5,50) -- (axis cs:2.5,1500);
\draw[racgrid,line width=0.4pt] (axis cs:5.5,50) -- (axis cs:5.5,1500);
\latencymodellabels

\nextgroupplot[
  ymin=100,
  ymax=3000
]
\addplot+[latency raw] coordinates {
  (0,143.5) (1,248.81) (2,857.32)
  (3,161.51) (4,166.18) (5,380.11)
  (6,379.07) (7,354.75) (8,1405.42)
};
\addplot+[latency rac] coordinates {
  (0,115.87) (1,128.08) (2,651.65)
  (3,165.74) (4,177.84) (5,235.19)
  (6,231.65) (7,262.28) (8,1124.74)
};
\addplot+[latency topk eight] coordinates {
  (0,193.1) (1,273.48) (2,965.09)
  (3,281.21) (4,340.11) (5,623.54)
  (6,460.5) (7,691.13) (8,2506.7)
};
\addplot+[latency topk four] coordinates {
  (0,194.2) (1,247.13) (2,715.85)
  (3,245.42) (4,269.47) (5,529.9)
  (6,351.49) (7,422.19) (8,1316.43)
};
\draw[racgrid,line width=0.4pt] (axis cs:2.5,100) -- (axis cs:2.5,3000);
\draw[racgrid,line width=0.4pt] (axis cs:5.5,100) -- (axis cs:5.5,3000);
\latencymodellabels

\end{groupplot}
\node[font=\racplottitlefont,anchor=north]
  at ($(group c1r1.south)+(0,-0.14cm)$) {(a) Mean TTFT};
\node[font=\racplottitlefont,anchor=north]
  at ($(group c2r1.south)+(0,-0.14cm)$) {(b) Median TTFT};
\node[font=\racplottitlefont,anchor=north]
  at ($(group c3r1.south)+(0,-0.14cm)$) {(c) 99th-percentile TTFT};
\node[font=\racplottitlefont,anchor=north]
  at ($(group c1r2.south)+(0,-1.72cm)$) {(d) Mean TPOT};
\node[font=\racplottitlefont,anchor=north]
  at ($(group c2r2.south)+(0,-1.72cm)$) {(e) Median TPOT};
\node[font=\racplottitlefont,anchor=north]
  at ($(group c3r2.south)+(0,-1.72cm)$) {(f) 99th-percentile TPOT};
\node[font=\racplotlabelfont,rotate=90]
  at ($(group c1r1.west)!0.5!(group c1r2.west)+(-0.78cm,0)$)
  {Latency (ms; log scale)};
\node[font=\racplotlabelfont]
  at ($(group c2r2.south)+(0,-1.40cm)$)
  {Decode bandwidth (Mbit/s; prefill bandwidth = 100 Mbit/s)};
\node
  at ($(group c2r1.north)+(0,0.78cm)$)
  {\ref{latencybarlegend}};
\end{tikzpicture}
\endgroup
\caption{End-to-end latency across models and bandwidth configurations. Each model group orders decode bandwidths as 10, 5, and 1 Mbit/s while the prefill bandwidth remains 100 Mbit/s. The six panels report the mean, median, and 99th percentile of time to first token (TTFT) and time per output token (TPOT) for Raw, RAC, TopK (8-bit ratio), and TopK (4-bit ratio). Vertical axes use logarithmic scales.}
\label{fig:performance-complete}
\end{figure*}

Fig.~\ref{fig:performance-complete} reports all three bandwidth configurations and all six latency statistics. Across nine model--link pairs, mean TTFT and TPOT are lower than Raw, with Raw-to-RAC ratios of 1.24--2.72$\times$ and 1.01--2.79$\times$; the 99th-percentile TTFT is also lower in all nine. At Qwen's 100/5-Mbit/s point, median TTFT, median TPOT, and 99th-percentile TPOT regress by 12.16, 6.81, and 11.66 ms, respectively; the last metric also regresses by 4.23 ms at 100/10 Mbit/s. Thus, lower mean latency does not eliminate median or tail variability. TPOT gains otherwise grow as decode bandwidth falls because serialization occupies more of the critical path. Because TopK has lower task quality (Table~\ref{tab:precision-complete}), its latency curves represent different quality operating points.

\subsection{Exploratory Joint Bandwidth Sensitivity}

\pgfplotsset{
  colormap={racblueheat}{
    rgb255=(247,251,255)
    rgb255=(222,235,247)
    rgb255=(198,219,239)
    rgb255=(158,202,225)
    rgb255=(107,174,214)
  },
  colormap={racorangeheat}{
    rgb255=(255,250,242)
    rgb255=(254,230,196)
    rgb255=(253,208,162)
    rgb255=(253,174,107)
    rgb255=(237,143,59)
  },
  colormap={racgreenheat}{
    rgb255=(247,252,245)
    rgb255=(229,245,224)
    rgb255=(199,233,192)
    rgb255=(161,217,155)
    rgb255=(116,196,118)
  }
}

\begin{figure*}[t]
\centering
\begin{tikzpicture}
\begin{groupplot}[
  rac data plot,
  group style={
    group size=3 by 1,
    horizontal sep=1.55cm
  },
  width=3.70cm,
  height=2.50cm,
  scale only axis,
  xmin=-0.5,
  xmax=4.5,
  ymin=-0.5,
  ymax=4.5,
  xtick={0,2,4},
  xticklabels={100,500,1000},
  ytick={0,1,2,3,4},
  yticklabels={1,5,10,25,75},
  tick align=outside,
  major tick length=1.2pt,
  tick label style={font=\racplottickfont},
  xlabel={prefill bandwidth (Mbit/s)},
  xlabel style={
    font=\racplotlabelfont,
    at={(axis description cs:0.5,-0.28)},
    anchor=north
  }
]

\nextgroupplot[
  colormap name=racblueheat,
  point meta min=0.9,
  point meta max=3.1,
  colorbar,
  colorbar style={
    width=0.80mm,
    height=2.50cm,
    at={(1.05,0.5)},
    anchor=west,
    ytick={1,2,3},
    tick label style={font=\racplottickfont},
    axis line style={draw=racgrid,line width=0.25pt},
    tick style={draw=racdark!70,line width=0.25pt},
    major tick length=1.0pt
  }
]
\addplot[
  matrix plot*,
  mesh/cols=5,
  point meta=explicit,
  draw=white,
  line width=0.25pt,
] coordinates {
  (0,0) [2.15]
  (1,0) [2.43]
  (2,0) [2.48]
  (3,0) [2.45]
  (4,0) [2.54]
  (0,1) [1.92]
  (1,1) [1.51]
  (2,1) [1.27]
  (3,1) [1.19]
  (4,1) [1.15]
  (0,2) [1.84]
  (1,2) [1.47]
  (2,2) [1.13]
  (3,2) [1.02]
  (4,2) [0.98]
  (0,3) [1.84]
  (1,3) [1.45]
  (2,3) [1.12]
  (3,3) [1.05]
  (4,3) [0.95]
  (0,4) [2.00]
  (1,4) [1.36]
  (2,4) [1.12]
  (3,4) [1.00]
  (4,4) [0.99]
};

\nextgroupplot[
  ytick=\empty,
  colormap name=racorangeheat,
  point meta min=0.9,
  point meta max=3.1,
  colorbar,
  colorbar style={
    width=0.80mm,
    height=2.50cm,
    at={(1.05,0.5)},
    anchor=west,
    ytick={1,2,3},
    tick label style={font=\racplottickfont},
    axis line style={draw=racgrid,line width=0.25pt},
    tick style={draw=racdark!70,line width=0.25pt},
    major tick length=1.0pt
  }
]
\addplot[
  matrix plot*,
  mesh/cols=5,
  point meta=explicit,
  draw=white,
  line width=0.25pt,
] coordinates {
  (0,0) [2.22]
  (1,0) [2.90]
  (2,0) [2.86]
  (3,0) [2.88]
  (4,0) [2.83]
  (0,1) [1.70]
  (1,1) [1.68]
  (2,1) [1.64]
  (3,1) [1.67]
  (4,1) [1.66]
  (0,2) [1.34]
  (1,2) [1.26]
  (2,2) [1.27]
  (3,2) [1.18]
  (4,2) [1.20]
  (0,3) [1.02]
  (1,3) [1.05]
  (2,3) [1.00]
  (3,3) [0.99]
  (4,3) [0.94]
  (0,4) [1.08]
  (1,4) [0.96]
  (2,4) [0.94]
  (3,4) [0.96]
  (4,4) [0.97]
};

\nextgroupplot[
  ytick=\empty,
  colormap name=racgreenheat,
  point meta min=0.9,
  point meta max=3.1,
  colorbar,
  colorbar style={
    width=0.80mm,
    height=2.50cm,
    at={(1.05,0.5)},
    anchor=west,
    ytick={1,2,3},
    tick label style={font=\racplottickfont},
    axis line style={draw=racgrid,line width=0.25pt},
    tick style={draw=racdark!70,line width=0.25pt},
    major tick length=1.0pt
  }
]
\addplot[
  matrix plot*,
  mesh/cols=5,
  point meta=explicit,
  draw=white,
  line width=0.25pt,
] coordinates {
  (0,0) [2.48]
  (1,0) [3.05]
  (2,0) [3.07]
  (3,0) [3.07]
  (4,0) [3.06]
  (0,1) [1.70]
  (1,1) [1.68]
  (2,1) [1.66]
  (3,1) [1.67]
  (4,1) [1.67]
  (0,2) [1.35]
  (1,2) [1.28]
  (2,2) [1.27]
  (3,2) [1.19]
  (4,2) [1.17]
  (0,3) [1.04]
  (1,3) [1.04]
  (2,3) [1.00]
  (3,3) [0.98]
  (4,3) [0.92]
  (0,4) [1.11]
  (1,4) [1.00]
  (2,4) [0.96]
  (3,4) [0.96]
  (4,4) [0.98]
};

\end{groupplot}

\node[
  anchor=north,
  font=\racplottitlefont
] at ($(group c1r1.south)+(0,-0.98cm)$) {(a) TTFT speedup};
\node[
  anchor=north,
  font=\racplottitlefont
] at ($(group c2r1.south)+(0,-0.98cm)$) {(b) TPOT speedup};
\node[
  anchor=north,
  font=\racplottitlefont
] at ($(group c3r1.south)+(0,-0.98cm)$) {(c) Throughput gain};
\node[
  rotate=90,
  font=\racplotlabelfont
] at ($(group c1r1.west)+(-0.82cm,0)$)
  {decode bandwidth (Mbit/s)};

\end{tikzpicture}
\caption{RAC improvement over Raw across the joint bandwidth sweep: (a) Raw/RAC mean-TTFT speedup, (b) Raw/RAC mean-TPOT speedup, and (c) RAC/Raw output-token-throughput gain. Each cell represents one prefill/decode bandwidth pair; values above one favor RAC. All panels use the same linear scale.}
\label{fig:sensitivity-complete}
\end{figure*}
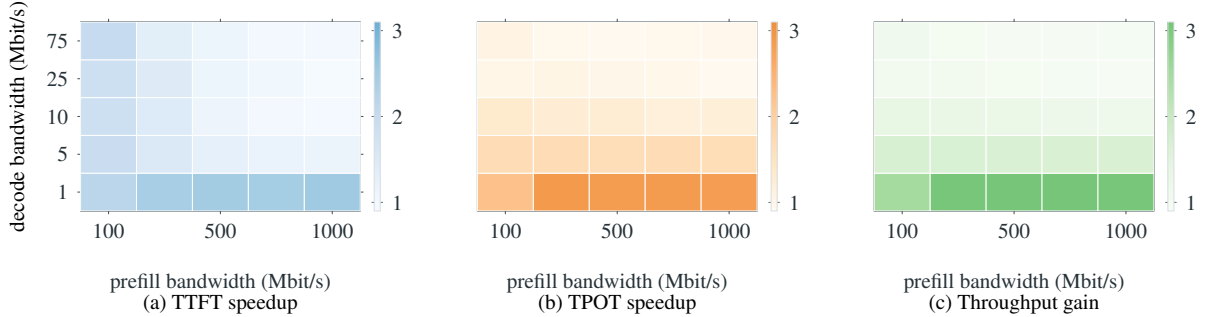

Because the model checkpoint identity was not retained, we interpret Fig.~\ref{fig:sensitivity-complete} only within the measured sweep. Across 25 bandwidth pairs, RAC has lower TTFT in 21, lower TPOT in 18, and higher throughput in 19. Its advantage is strongest at lower decode bandwidths, while several high-bandwidth pairs cross over in favor of Raw. The measured crossover shows that serialization savings must exceed codec and fixed transfer costs and that the calibrated operating point should reflect the target network regime.

\subsection{Component Ablation}

\begin{table*}[!t]
\centering
\caption{Quality ablation across the three evaluated models. Accuracy, exact-match, and MathVerify values use a 0--100 percentage scale. Boldface and underlining mark the best and second-best values among RAC and its ablation variants, respectively.}
\label{tab:ablation-complete}
\begingroup
\footnotesize
\sisetup{
  mode=math,
  group-digits=none,
  detect-weight=true,
  table-number-alignment=center,
  table-text-alignment=center
}
\setlength{\tabcolsep}{2.2pt}
\renewcommand{\arraystretch}{1.00}
\begin{tabular}{@{}
  l
  l
  @{\hspace{6pt}}
  *{5}{S[table-format=2.2]}
  @{}}
\toprule
\multirow{2}{*}{\textbf{Model}} &
\multirow{2}{*}{\textbf{Variant}} &
  \mbox{\textbf{Language modeling}} &
  {\textbf{Commonsense}} &
  \multicolumn{3}{c}{\textbf{Mathematical reasoning}} \\
\cmidrule(lr){3-3}\cmidrule(lr){4-4}\cmidrule(lr){5-7}
& &
  {\shortstack{\textbf{WikiText-2}\\PPL $\downarrow$}} &
  {\shortstack{\textbf{HellaSwag}\\Norm.\ Acc.\ $\uparrow$}} &
  {\shortstack{\textbf{GSM8K}\\Flexible EM $\uparrow$}} &
  {\shortstack{\textbf{GSM8K}\\Strict EM $\uparrow$}} &
  {\shortstack{\textbf{MATH-500}\\MathVerify $\uparrow$}} \\
\midrule
\multirow{4}{*}{\shortstack[l]{\textbf{GLM-4}\\9B-0414}} & \textbf{\system (Ours)}
  & {\underline{39.66}} &
    {\underline{80.10}} &
    {\textbf{80.21}} &
    {\textbf{80.36}} &
    {\textbf{56.60}} \\
\cmidrule(l){2-7}
& w/o prediction
  & {\underline{39.66}} &
    {\textbf{80.16}} &
    69.83 & 70.51 & 34.60 \\
& w/o search
  & {\textbf{38.86}} &
    80.04 & 77.71 & 78.39 & 47.40 \\
& w/o affine alignment
  & 39.82 & 80.07 &
    {\underline{78.92}} &
    {\underline{79.38}} & {\underline{49.40}} \\

\midrule
\multirow{4}{*}{\shortstack[l]{\textbf{Qwen3}\\30B-A3B}} & \textbf{\system (Ours)}
  & {\textbf{19.45}} &
    77.42 & {\textbf{87.11}} &
    {\textbf{89.76}} &
    {\textbf{79.80}} \\
\cmidrule(l){2-7}
& w/o prediction
  & {\textbf{19.45}} &
    {\textbf{77.80}} &
    {\underline{86.08}} &
    88.63 & 78.00 \\
& w/o search
  & 19.49 & 77.38 & 84.31 & 88.55 & {\underline{78.60}} \\
& w/o affine alignment
  & {\underline{19.48}} &
    {\underline{77.45}} &
    86.02 & {\underline{88.86}} & {\underline{78.60}} \\

\midrule
\multirow{4}{*}{\shortstack[l]{\textbf{Llama-3.3}\\70B-Instruct}} & \textbf{\system (Ours)}
  & {\textbf{13.54}} &
    {\underline{84.87}} &
    {\textbf{92.04}} &
    {\textbf{76.27}} &
    {\textbf{65.40}} \\
\cmidrule(l){2-7}
& w/o prediction
  & {\textbf{13.54}} &
    {\textbf{84.88}} &
    {\underline{91.89}} &
    71.27 & 61.20 \\
& w/o search
  & {\underline{13.56}} &
    84.81 & 91.43 &
    {\underline{75.65}} & 62.00 \\
& w/o affine alignment
  & 13.59 & 84.83 & 91.51 & 74.75 & {\underline{63.80}} \\
\bottomrule
\end{tabular}
\endgroup
\end{table*}

Table~\ref{tab:ablation-complete} compares reference-aware compression with configurations that remove prediction, search, or affine alignment. All three reduce both GSM8K exact-match scores and MATH-500 accuracy for every model, whereas WikiText-2 and HellaSwag differences are small and mixed. For GLM, removing prediction changes GSM8K flexible and strict exact match by $-10.38$ and $-9.85$ points and MATH-500 by $-22.0$ points. Prediction removal also lowers strict exact match by 1.13 points for Qwen and 5.00 points for Llama. These results show that prediction contributes temporal reference structure; search supplies historical structure for prefill uplinks, and affine alignment adapts references across the referenced prefill and decode transfers.

\subsection{Runtime Overhead}
\label{sec:runtime-overhead}

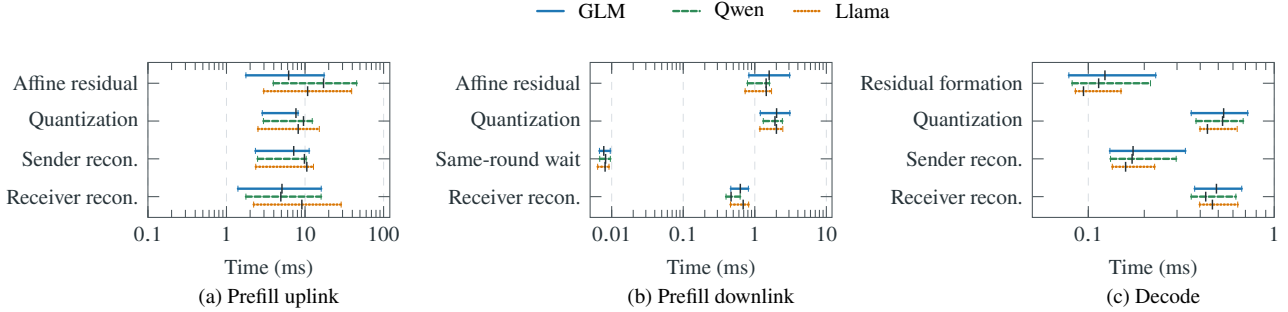
\begin{figure*}[t]
\centering
\begingroup
\tikzset{
  overhead-glm/.style={rac model glm},
  overhead-qwen/.style={rac model qwen},
  overhead-llama/.style={rac model llama}
}
\newcommand{\OverheadRange}[5]{%
  \draw[overhead-#1,line width=0.90pt,line cap=round]
    (axis cs:#3,#2) -- (axis cs:#5,#2);
  \draw[overhead-#1,solid,line width=0.35pt]
    (axis cs:#3,{#2-0.08}) -- (axis cs:#3,{#2+0.08});
  \draw[overhead-#1,solid,line width=0.35pt]
    (axis cs:#5,{#2-0.08}) -- (axis cs:#5,{#2+0.08});
  \draw[draw=racdark,line width=0.55pt]
    (axis cs:#4,{#2-0.13}) -- (axis cs:#4,{#2+0.13});
}
\begin{tikzpicture}
\begin{groupplot}[
  rac data plot,
  group style={group size=3 by 1,horizontal sep=2.65cm},
  width=3.20cm,
  height=2.05cm,
  scale only axis,
  xmode=log,
  xmajorgrids,
  log ticks with fixed point,
  clip=false,
  xlabel={Time (ms)},
]
\nextgroupplot[
  xmin=0.1,xmax=120,xtick={0.1,1,10,100},
  ymin=-0.55,ymax=3.55,
  ytick={3,2,1,0},
  yticklabels={{Affine residual},{Quantization},{Sender recon.},{Receiver recon.}},
  y tick label style={font=\racplottickfont,align=right},
]

\OverheadRange{glm}{3.21}{1.767581}{6.182809}{17.46643}
\OverheadRange{qwen}{3.00}{3.947858}{17.202125}{45.386895}
\OverheadRange{llama}{2.79}{2.977442}{10.785383}{39.21963}
\OverheadRange{glm}{2.21}{2.856588}{7.661568}{8.146944}
\OverheadRange{qwen}{2.00}{2.954235}{9.557862}{12.351601}
\OverheadRange{llama}{1.79}{2.516065}{8.184832}{15.198208}
\OverheadRange{glm}{1.21}{2.335638}{7.173939}{11.270308}
\OverheadRange{qwen}{1.00}{2.488082}{9.821389}{10.347827}
\OverheadRange{llama}{0.79}{2.357836}{10.5344}{12.811776}
\OverheadRange{glm}{0.21}{1.398905}{5.086208}{16.068608}
\OverheadRange{qwen}{0.00}{1.767824}{4.919347}{16.007785}
\OverheadRange{llama}{-0.21}{2.217057}{9.107456}{28.817408}

\nextgroupplot[
  xmin=0.005,xmax=12,xtick={0.01,0.1,1,10},
  ymin=-0.55,ymax=3.55,
  ytick={3,2,1,0},
  yticklabels={{Affine residual},{Quantization},{Same-round wait},{Receiver recon.}},
  y tick label style={font=\racplottickfont,align=right},
]

\OverheadRange{glm}{3.21}{0.827368}{1.58817}{3.067595}
\OverheadRange{qwen}{3.00}{0.790064}{1.439938}{1.601084}
\OverheadRange{llama}{2.79}{0.733289}{1.445824}{1.7152}
\OverheadRange{glm}{2.21}{1.194752}{2.016973}{3.084902}
\OverheadRange{qwen}{2.00}{1.315069}{1.914726}{2.429491}
\OverheadRange{llama}{1.79}{1.178115}{2.004992}{2.46272}
\OverheadRange{glm}{1.21}{0.006749}{0.007757}{0.009552}
\OverheadRange{qwen}{1.00}{0.00681}{0.008221}{0.009589}
\OverheadRange{llama}{0.79}{0.006367}{0.008093}{0.00917}
\OverheadRange{glm}{0.21}{0.462044}{0.627456}{0.81367}
\OverheadRange{qwen}{0.00}{0.396583}{0.466074}{0.623903}
\OverheadRange{llama}{-0.21}{0.457769}{0.688128}{0.820224}

\nextgroupplot[
  xmin=0.05,xmax=1,xtick={0.1,1},
  ymin=-0.55,ymax=3.55,
  ytick={3,2,1,0},
  yticklabels={{Residual formation},{Quantization},{Sender recon.},{Receiver recon.}},
  y tick label style={font=\racplottickfont,align=right},
]

\OverheadRange{glm}{3.21}{0.078613}{0.12288}{0.2304}
\OverheadRange{qwen}{3.00}{0.081874}{0.113664}{0.216064}
\OverheadRange{llama}{2.79}{0.085227}{0.094208}{0.149955}
\OverheadRange{glm}{2.21}{0.357889}{0.534528}{0.719749}
\OverheadRange{qwen}{2.00}{0.379805}{0.52736}{0.680694}
\OverheadRange{llama}{1.79}{0.398945}{0.437248}{0.632259}
\OverheadRange{glm}{1.21}{0.130664}{0.17408}{0.3328}
\OverheadRange{qwen}{1.00}{0.131711}{0.172032}{0.29696}
\OverheadRange{llama}{0.79}{0.134627}{0.15872}{0.227328}
\OverheadRange{glm}{0.21}{0.373356}{0.488858}{0.668672}
\OverheadRange{qwen}{0.00}{0.35789}{0.428032}{0.621947}
\OverheadRange{llama}{-0.21}{0.397297}{0.465101}{0.636355}
\end{groupplot}
\node[anchor=north,font=\racplottitlefont] at
  ($(group c1r1.south)+(0,-0.82cm)$) {(a) Prefill uplink};
\node[anchor=north,font=\racplottitlefont] at
  ($(group c2r1.south)+(0,-0.82cm)$) {(b) Prefill downlink};
\node[anchor=north,font=\racplottitlefont] at
  ($(group c3r1.south)+(0,-0.82cm)$) {(c) Decode};
\draw[overhead-glm,line width=0.90pt,line cap=round]
  ($(group c2r1.north)+(-2.30cm,0.68cm)$) -- ++(0.35cm,0);
\node[anchor=west,font=\racplotlegendfont] at
  ($(group c2r1.north)+(-1.88cm,0.68cm)$) {GLM};
\draw[overhead-qwen,line width=0.90pt,line cap=round]
  ($(group c2r1.north)+(-0.50cm,0.68cm)$) -- ++(0.35cm,0);
\node[anchor=west,font=\racplotlegendfont] at
  ($(group c2r1.north)+(-0.08cm,0.68cm)$) {Qwen};
\draw[overhead-llama,line width=0.90pt,line cap=round]
  ($(group c2r1.north)+(1.10cm,0.68cm)$) -- ++(0.35cm,0);
\node[anchor=west,font=\racplotlegendfont] at
  ($(group c2r1.north)+(1.52cm,0.68cm)$) {Llama};
\end{tikzpicture}
\endgroup
\caption{Runtime summaries for selected post-activation stages: (a) prefill uplink, (b) prefill downlink, and (c) decode. Segments span the mean to the 99th percentile; dark ticks mark the 95th percentile. All horizontal axes are logarithmic.}
\label{fig:overhead-percentile-ranges}
\end{figure*}

Fig.~\ref{fig:overhead-percentile-ranges} reports selected post-activation stages for each transfer. Affine residual construction has the largest prefill-uplink 99th percentile, while every displayed decode stage has a sub-millisecond mean, consistent with full-sequence versus one-token processing. The mean same-round wait remains below 0.007 ms. Token hashing, span selection, reference prefetch, predictor execution, and predictor wait are omitted because they overlap model computation or communication. The figure therefore focuses on stages exposed after the activation becomes available.

\section{Related Work}

\textbf{Split inference and LLM serving.} Device--cloud systems partition consecutive submodels and optimize placement under latency, accuracy, and resource constraints~\cite{kang2017neurosurgeon,eshatifar2021jointdnn,li2018jalad,borzunov2023petals,ye2025jupiter}. RAC leaves the partitioned LLM unchanged, calibrating the boundary codec offline. LLM-serving systems optimize batching, attention-memory management, phase scheduling, and prefill--decode disaggregation~\cite{yu2022orca,kwon2023vllm,agrawal2024sarathi,zhong2024distserve,patel2024splitwise}. State offloading, prompt reuse, and KV-cache transport instead move model state or persistent attention history~\cite{sheng2023flexgen,gim2024promptcache,liu2024cachegen}. RAC operates under fixed three-way placement and compresses transient boundary activations without changing phase placement or transporting the KV cache.

\textbf{Intermediate-representation compression.} Learned bottlenecks train compact transmitted features~\cite{eshatifar2019bottlenet,shao2020bottlenetpp,matsubara2022supervised,matsubara2023sc2}, whereas quantization, sparsification, mixed precision, and smoothing encode existing tensors without model changes~\cite{jacob2018quant,dettmers2022llmint8,yao2022zeroquant,xiao2023smoothquant}. Direct quantization covers the full tensor range; RAC quantizes only the aligned difference. RAC constructs phase-specific references from repeated spans, same-round boundaries, or consecutive decode steps, then calibrates grouping, bit width, and outlier handling through sender-side wire-format reconstruction and packed-cost accounting.

\textbf{Privacy-oriented split inference.} Perturbation and pruning defenses reduce information retained by transmitted features~\cite{mireshghallah2020shredder,ding2024patrol}. RAC targets communication efficiency rather than formal privacy and can be combined with such defenses; compression alone is not a confidentiality guarantee.

\section{Conclusion}

We presented RAC, a reference-aware boundary codec for split LLM inference. RAC retrieves exact-token historical spans for prefill uplinks, reuses the reconstructed uplink state for same-round prefill downlinks, and generates boundary-specific decode references using lightweight causal predictors. For each referenced transfer, RAC fits a grouped affine mapping, quantizes the aligned residual using calibrated 4-bit or 8-bit groups with optional prefill outliers, and reconstructs the transmitted activation at the sender to keep subsequent references synchronized with the receiver. Its activation payload is approximately $27.8\%$ of Raw for prefill and $25.0\%$ for decode. Across nine model--link pairs, Raw-to-RAC mean TTFT and TPOT ratios are 1.24--2.72$\times$ and 1.01--2.79$\times$, while the 12 non-perplexity score changes range from $-0.40$ to $+2.50$ points.

\clearpage
\bibliographystyle{IEEEtran}
\bibliography{references}

\end{document}